\documentclass[pra,twocolumn,superscriptaddress]{revtex4-2}

\usepackage[usenames, dvipsnames]{xcolor}
\usepackage[utf8]{inputenc}
\usepackage{hyperref}
\hypersetup{colorlinks,allcolors=blue}
\usepackage{amsmath,revsymb,amssymb}
\usepackage{graphicx}
\usepackage{xcolor}
\usepackage{bbold,comment}
\usepackage{braket,ulem}
\usepackage{booktabs}
\usepackage{float}
\usepackage{physics}
\usepackage{multirow}
\usepackage{tabularx}
\usepackage{array}
\newcolumntype{C}[1]{>{\centering\let\newline\\\arraybackslash\hspace{0pt}}m{#1}}
\newcolumntype{Y}{>{\centering\arraybackslash}X}

\usepackage{color, colortbl}
\definecolor{Gray}{gray}{0.9}
\newcolumntype{g}{>{\columncolor{Gray}}l}

\renewcommand{\emph}{\textit}

\begin{document}
    
\title{Experimental test of sequential weak measurements for certified quantum randomness extraction}

% Authors
\author{Giulio Foletto}
\thanks{These authors contributed equally to this work.}
\affiliation{Dipartimento di Ingegneria dell'Informazione, Universit\`a degli Studi di Padova, via Gradenigo 6B, IT-35131 Padova, Italy}

\author{Matteo Padovan}
\thanks{These authors contributed equally to this work.}
\affiliation{Dipartimento di Ingegneria dell'Informazione, Universit\`a degli Studi di Padova, via Gradenigo 6B, IT-35131 Padova, Italy}

\author{Marco Avesani}
\affiliation{Dipartimento di Ingegneria dell'Informazione, Universit\`a degli Studi di Padova, via Gradenigo 6B, IT-35131 Padova, Italy}

\author{Hamid Tebyanian}
\affiliation{Dipartimento di Ingegneria dell'Informazione, Universit\`a degli Studi di Padova, via Gradenigo 6B, IT-35131 Padova, Italy}

\author{Paolo Villoresi}
\affiliation{Dipartimento di Ingegneria dell'Informazione, Universit\`a degli Studi di Padova, via Gradenigo 6B, IT-35131 Padova, Italy}
\affiliation{Padua Quantum Technologies Research Center, Universit\`a degli Studi di Padova, via Gradenigo 6B, IT-35131 Padova, Italy}

\author{Giuseppe Vallone}
\email{vallone@dei.unipd.it}
\affiliation{Dipartimento di Ingegneria dell'Informazione, Universit\`a degli Studi di Padova, via Gradenigo 6B, IT-35131 Padova, Italy}
\affiliation{Dipartimento di Fisica e Astronomia, Universit\`a degli Studi di Padova, via Marzolo 8, IT-35131 Padova, Italy}
\affiliation{Padua Quantum Technologies Research Center, Universit\`a degli Studi di Padova, via Gradenigo 6B, IT-35131 Padova, Italy}
\begin{abstract}
    Quantum nonlocality offers a secure way to produce random numbers: Their unpredictability is intrinsic and can be certified just by observing the statistic of the measurement outcomes, without assumptions on how they are produced.
    To do this, entangled pairs are generated and measured to violate a Bell inequality with the outcome statistics.
    However, after a projective quantum measurement, entanglement is entirely destroyed and cannot be used again.
    This fact poses an upper bound to the amount of randomness that can be produced from each quantum state when projective measurements are employed.
    Instead, by using weak measurements, some entanglement can be maintained and reutilized, and a sequence of weak measurements can extract an unbounded amount of randomness from a single state as predicted in Phys. Rev. A \textbf{95}, 020102(R) (2017).
    We study the feasibility of these weak measurements, analyze the robustness to imperfections in the quantum state they are applied to, and then test them using an optical setup based on polarization-entangled photon pairs. 
    We show that the weak measurements are realizable, but can improve the performance of randomness generation only in close-to-ideal conditions.
\end{abstract}

\maketitle

% Beginning of main text
\section{INTRODUCTION}
Classical random number generators cannot produce genuine randomness as they rely on algorithms or deterministic phenomena.
However, quantum physics offers several solutions for producing secure and private random numbers \cite{Ma2016,Herrero-Collantes2017}.
The simplest one arises from the superposition principle, which makes quantum measurements probabilistic on most states.
This idea can be exploited, for example, by identifying two mutually unbiased bases, preparing a physical system in a state belonging to one, and measuring it with respect to the other.
Neglecting experimental imperfections, in the long run, the measurement outcomes will be random and uniformly distributed. 
However, this relies on at least two strong assumptions: knowledge of the quantum state and accurate control of the measurement being made.
These are often hard to verify in practice, and hence leave an opening for potential attacks.

A different approach to quantum randomness starts from the concept of nonlocality \cite{Brunner2014}, for which the outcomes of measurements on some multipartite systems generate correlations that cannot be explained by theories that are local and realistic.
In arguably one of the most important results of quantum theory, J. S. Bell showed that there are relations between the statistics of the measurement outcomes that must hold for such theories but are violated by quantum physics \cite{Bell1964}.
These relations, now called Bell inequalities, have been violated experimentally countless times, thus proving that local realistic theories are incompatible with the experimental data~\cite{Aspect1981,Aspect1982a,Aspect1982b,Hensen2015,Giustina2015,Shalm2015}.
Assuming that no-signaling still holds, the measurements that violate the inequalities are intrinsically unpredictable, and hence they can produce random numbers~\cite{Colbeck2006}.
Protocols that exploit the violation of a Bell inequality are often termed \textit{device independent}, because this violation does not require any assumption on the nature of the state nor the measurements, and hence is independent of the inner workings of the devices in use.
This level of security is higher than that of other frameworks (e.g., trusted device \cite{Herrero-Collantes2017} or semi-device-independent \cite{Li2011,Nie2016,Avesani2018,Avesani2021}), which require full or partial trust on the devices and cannot allow them to be controlled by an adversary, something that, instead, is tolerated in the device-independent case.

This abstract intuition has been made more quantitative with the study and development of device-independent random number generators \cite{Pironio2010,Bierhorst2018,Liu2018,Shen2018,Zhang2020,Shalm2021,Liu2021}.
In general, these instruments consist of a source of entangled states, which are necessary for violating a Bell inequality, and some measurement stations that receive each subsystem, measure it, and attempt to observe nonlocality using the result statistics. 
The amount of randomness that can be extracted from the measurement outcomes depends on the strength of the violation.
For instance, most implementations use two-qubit states, such as polarization-entangled photon pairs and exploit the CHSH inequality~\cite{Clauser1969}.
A limitation of this scheme is that the projective measurements irreversibly destroy entanglement, hence each pair can contribute to only one violation and produce at most one bit of randomness if the outcomes on one subsystem are used, or 1.23 bits if both parties are considered \cite{Pironio2010}.

Although there are other ways to overcome this bound (see, e.g., Ref. \cite{Andersson2018}), we shall focus on the use of the weak measurement~\cite{Aharonov1988}.
Throughout the last three decades, this tool has found many diverse applications, from the amplification of feeble quantities \cite{Hosten2008, Dixon2009, Jayaswal2014}, to the measurement of incompatible observables \cite{Kocsis2011, Piacentini2016, Chen2019}, through quantum state reconstruction \cite{Lundeen2011, Vallone2016, Thekkadath2016, Calderaro2018}.
Recently, it has been exploited for sequential protocols, in which a system undergoes multiple measurements without ever completely collapsing or losing its useful quantum features, which can be harvested repeatedly.
In this manner, Bell inequalities can be violated more times \cite{Schiavon2016, Avella2017, Tavakoli2018, Hu2018, Foletto2020}, quantum random access codes can be used by two parties \cite{Mohan2019, Foletto2020b}, and quantum instruments can be tested \cite{Anwer2020}.
More important for this work is using weak measurements to produce random bits from the same physical system repeatedly \cite{Curchod2017a, Coyle2018}.
The authors of Ref. \cite{Curchod2017a} proposed a device-independent protocol based on the sequential violation of a CHSH-like inequality on a bipartite entangled state.
In the case of perfect state preparation and an infinite sequence of ideal measurements, their scheme can produce an unbounded amount of random bits from the outcomes of local measurements on one subsystem.

Albeit valid only in an ideal scenario, this fact encourages an evaluation of the practical feasibility of this protocol, which is the aim of this work.
We first theoretically analyze its robustness to imperfections and then show a proof-of-concept experimental implementation based on bulk polarization optics that highlights the difficulties inherent in these measurements, but can be a starting point for further developments using setups of higher accuracy. 

\section{THEORETICAL MODEL}
\label{sec:model}
In Secs. \ref{sub-sec:sequence}, \ref{sub-sec:extraction}, and \ref{sub-sec:non_maximal_violations}, we summarize and extensively comment on the protocol proposed in Ref. \cite{Curchod2017a} to set the framework we are working in and explain the notation.
Then, in Sec. \ref{sub-sec:robustness}, we present the main theoretical result of this work.
We first introduce a simple model to characterize the robustness of the protocol to experimental imperfections, and then apply it in numerical simulations to study how much randomness can be generated under different noise conditions.

\subsection{The sequential measurement protocol}
\label{sub-sec:sequence}
\begin{figure}
    \centering
    \includegraphics[width=\linewidth]{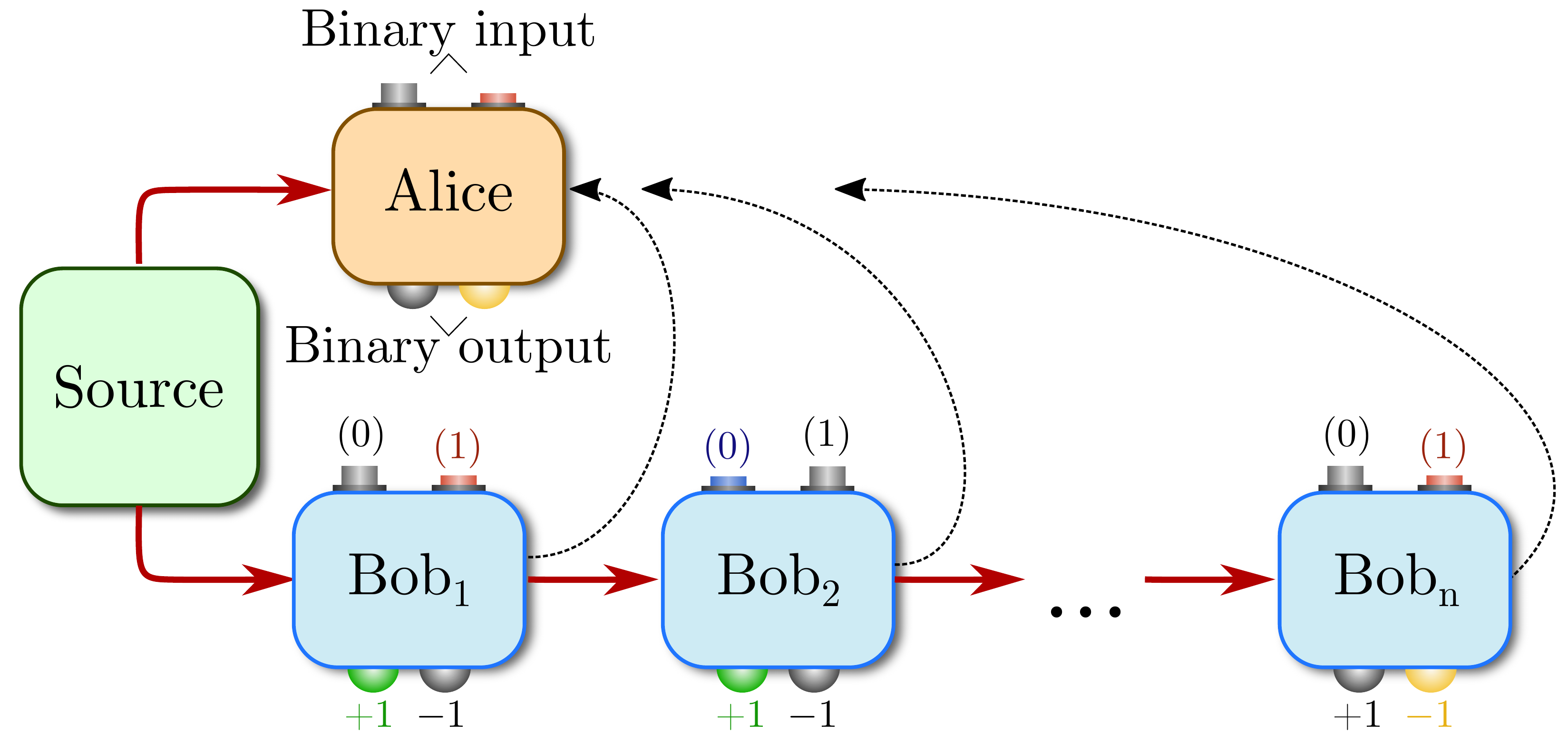}\\
    \caption{Scheme of the sequential protocol.
    The dotted black line indicates that, before attempting to violate the Bell inequality with a Bob, Alice requires the history of outcomes of the previous observers.
    This communication can also be replaced by post-selection.
    Alice's measurements are described by Eq. \eqref{eq:alice_observables}, whereas the Bobs' are in Eq. \eqref{eq:bob_observables}.}
    \label{fig:ProtocolScheme}
\end{figure}
The protocol, schematically depicted in Fig. \ref{fig:ProtocolScheme}, starts with a pure two-qubit entangled state, shared between two parties called Alice and ${\rm Bob}_1$:
\begin{equation}
    \ket{\psi_1} = \cos \theta_1 \ket{00} + \sin \theta_1 \ket{11},
    \label{eq:psi_1}
\end{equation}
where $\theta_1 \in [0, \pi/4]$ is an indicator of the amount of entanglement in the state, indeed if $\theta_1 = 0$, $\ket{\psi_1}$ is separable, whereas if $\theta_1 = \pi/4$, the state is a maximally entangled Bell pair.
In the best-case scenario, the protocol starts with $\theta_1 = \pi/4$ and hence can achieve the best performance.
We now focus on the action of ${\rm Bob}_1$, who is the first in a sequence of observers that work on the same half of the entangled pair.
We will describe Alice's role in Sec. \ref{sub-sec:extraction}.
${\rm Bob}_1$ selects one of two observables to be measured: 
\begin{equation}
\label{eq:bob_observables}
\begin{aligned}
    B^{(0)} &= \sigma_z, \\
    B^{(1)}_{1} &= \cos \qty(2\xi_1)\sigma_x,
\end{aligned}
\end{equation}
in which $\xi_1 \in [0, \pi/4]$ quantifies the strength of the measurement of $\sigma_x$, which is projective for $\xi_1 = 0$, completely noninteractive for $\xi_1 = \pi/4$, and generically weak for any value in between.
In this way, the observable $B^{(1)}_1$ corresponds to a generalized measurement of $\sigma_x$ with Kraus operators $K_{1,\pm}=\frac12[(\cos\xi_1+\sin\xi_1)\openone\pm (\cos\xi_1-\sin\xi_1)\sigma_x]$.
Both observables return binary outcomes $y_1 = \pm 1$, collected by ${\rm Bob}_1$.
The state of Eq. \eqref{eq:psi_1} is balanced with respect to the measurement of $B^{(1)}_1$, meaning that both outcomes can happen with $\frac12$ probability.
These outcomes are used to generate random bits, whereas those of $B^{(0)}$ are needed to violate a Bell inequality and certify that state and measurements are indeed those we are describing.

After the measurement, the state takes the form, 
\begin{equation}
\ket{\psi_{2,\vec{h}}} = U_{A,2,\vec{h}}\otimes U_{B,2,\vec{h}} (\cos(\theta_2)\ket{00} + \sin(\theta_2)\ket{11}),
\end{equation}
where the unitary operations $U_{A,2,\vec{h}}, U_{B,2,\vec{h}}$ depend on ${\rm Bob}_1$'s strength and outcome.
We use symbol $\vec{h}$ to label the \textit{history} of outcomes at the past steps (which in this case contains only one datum, $y_1$).
The value of the parameter $\theta_2$ does not depend on ${\rm Bob}_1$'s outcome but still depends on his strength, although we do not highlight this in the notation.

Using his knowledge of the outcome, ${\rm Bob}_1$ applies $U^\dagger_{B,2,\vec{h}}$ to his local state and sends it to ${\rm Bob}_2$, who can proceed with a similar step.
The purpose of this action is to make ${\rm Bob}_2$'s state again balanced with respect to the outcomes of $B^{(1)}$, so that he can again produce random bits by measuring the $B^{(1)}$ observable.
The protocol can continue with an unlimited sequence of measurements of $B^{(1)}$ or can be stopped with a strong measurement of $B^{(0)}$. 
There is no reason to continue the protocol after this because the post-measurement state is no longer entangled.
To make a long sequence of measurements, we can imagine that $B^{(0)}$ is chosen with low probability, just enough to provide the statistics that violate a Bell inequality.

\subsection{Extraction of random bits}
\label{sub-sec:extraction}
We now move to the other half of the entangled pair, held by Alice.
The outcomes of her measurements do not directly provide random bits, but are correlated with those of the Bobs to violate a Bell inequality.
For each quantum system, Alice selects a step $k$ and one of the two projective observables:
\begin{equation}
\label{eq:alice_observables}
    \begin{aligned}
    A^{(0)}_k &= \cos(\mu_k)\sigma_z + \sin(\mu_k)\sigma_x, \\
    A^{(1)}_k &= \cos(\mu_k)\sigma_z - \sin(\mu_k)\sigma_x,
    \end{aligned}
\end{equation}
where $\mu_k = \arctan(\sin(2\theta_k))$.
These choices change for each quantum state and the Bobs must not know them until after all actions and measurements are completed.
Before measuring her selected observable, Alice applies the unitary transformation $U_{A,k,\vec{h}}^\dagger$, so that again the global state takes the form of Eq. \eqref{eq:psi_1} (with the generic angle $\theta_k$).
To do this, she must wait for ${\rm Bob}_1,\cdots,{\rm Bob}_{k-1}$ to measure their half of the state and to send her their history $\vec{h}$ of outcomes.
However, she is careful to measure her state outside of the light cone of ${\rm Bob}_k$'s basis choice, otherwise the estimation of the Bell quantity would be affected by the locality loophole \cite{Larsson2014}.

By correlating their outcomes on multiple statistical repetition of this test, Alice and ${\rm Bob}_k$ can compute the quantity
\begin{equation}
    \begin{split}
    I_{k} = \beta_k \expval{B^{(0)}} &+ 
        \expval{A^{(0)}_kB^{(0)}} +
        \expval{A^{(0)}_kB^{(1)}_k} \\ &+
        \expval{A^{(1)}_kB^{(0)}} -
        \expval{A^{(1)}_kB^{(1)}_k},
    \end{split}
    \label{eq:bellQuantity}
\end{equation}
where $\beta_k = 2\cos(2\theta_k)/\sqrt{1+\sin^2(2\theta_k)}$, and the brackets denote the expectation value.
A local-hidden-variables model would restrict $I_k$ with the CHSH-like Bell inequality $I_k \leq \beta_k+2$, however, quantum theory allows a larger upper bound \cite{Acin2012}:
\begin{equation}
    I_{\mathrm{max},k} = \sqrt{2\qty(4+\beta_k^2)}.
\end{equation}
Crucially, observing this maximal value certifies that the state is the one described by Eq. \eqref{eq:psi_1} and the measurements are those of Eqs. \eqref{eq:bob_observables}, \eqref{eq:alice_observables}, with $\xi_k=0$, because this is the \emph{only} configuration (up to unitary transformations) that can reach this upper bound.
Moreover, since $\expval{B^{(1)}_k}=0$ on this state, this also certifies that the outcomes of $B^{(1)}_k$ are uniformly distributed and private, and therefore can be used as random bits, the ultimate goal of this scheme.
This is because of the monogamy of entanglement, which is reflected by the fact that the state \eqref{eq:psi_1} is bipartite and pure, and hence cannot be correlated with any information held by a third party.

The certification does not apply only to the outcomes of $B^{(1)}_k$ that are used in the estimation of $I_k$, but also to all the other outcomes of the same measurement and with the same history $\vec{h}$ generated by other entangled pairs (regardless of Alice's actions for those pairs).
Indeed, the Bobs' devices cannot know a priori which step $k$ will be chosen for each entangled pair, and therefore cannot apply different strategies to the quantum systems.
Any attempt to cheat is detected in the estimation of $I_k$, although only a subset of the outcomes contributes to said estimation.
Similarly, each entangled pair produces many outcomes, one at each step, and only one of these contributes to the estimation of the Bell quantity, the one corresponding to the step $k$ chosen by Alice for this pair.
This does not mean that the others are useless: after sufficiently many runs of the experiment, all the steps and all the possible histories undergo the certification and the randomness of all these outcomes is validated.

As previously mentioned, Alice requires to know the history $\vec{h}$ of ${\rm Bob}_1,\cdots,{\rm Bob}_{k-1}$'s outcomes to apply $U_{A,k,\vec{h}}^\dagger$.
This is an important limitation to the practicality of this scheme.
Indeed, the Bobs need a fast communication channel to send their outcomes to Alice just after they have produced them, so that she can apply the unitary transformation and measure her state before entering the light cone of ${\rm Bob}_k$'s basis choice, otherwise she would open the locality loophole.
If this communication is deceitful or contains error, the amount of certifiable randomness is reduced because Alice applies the wrong unitary transformation (but this is a denial of service, not a security risk).

A probably easier alternative is that Alice randomly chooses among the $2^{k-1}$ possible histories and learns whether her guess was correct only afterward, without the need for communication between the measurements.
Whenever her guess is wrong, she and ${\rm Bob}_k$ do not use their outcomes in the estimation of the Bell quantity, but none of the bits generated by the Bobs are thrown away: they will be certified later when Alice guesses correctly.
Just as above, even if the Bobs' devices are dishonest, they cannot predict Alice's chosen history, therefore they cannot apply different strategies to different entangled pairs and make those for which Alice's guess is wrong less secure.
This strategy introduces a larger delay between the production of the outcomes and their certification, because many more runs are needed to achieve the necessary statistics.
Yet, it does not decrease the randomness generation rate, because once the certification is done, it applies to all outcomes, not only to the few that were generated when Alice guessed right.
The \textit{net} generation rate is still reduced because Alice spends randomness in the choice of history, but this cost can be lessened using an unbalanced distribution.

Like in event-ready Bell tests \cite{Hensen2015}, discarding outcomes from the certification does not open any loophole \cite{Larsson2014}.
Indeed, the application of the random transformation $U^\dag_{A,k,\vec h}$ followed by one of the two measurements $A^{(0)}_k$ or $A^{(1)}_k$, can be interpreted as a measurement randomly chosen between $2^k$ different observables (defined as $U_{A,k,\vec h}A^{(0)}_kU^\dag_{A,k,\vec h}$ and $U_{A,k,\vec h}A^{(1)}_kU^\dag_{A,k,\vec h}$ with the $2^{k-1}$ possible choices of $\vec h$).
The actual history of outcomes provided to ${\rm Bob}_1,\cdots,{\rm Bob}_{k-1}$ by their devices plays the role of a description of the prepared entangled state shared between Alice and ${\rm Bob}_k$, given by $\ket{\psi_{k,\vec h}}=U_{A,k,\vec h}\otimes \openone_B (\cos(\theta_{k})\ket{00}+\sin(\theta_{k})\ket{11})$.
Therefore, this scenario can be seen as a Bell test where Alice can choose between $2^{k}$ observables and ${\rm Bob}_k$ can choose between two observables, corresponding to $2^{k-1}$ Bell inequalities.
Depending on the prepared state $\ket{\psi_{k,\vec h}}$, only one of the $2^{k-1}$ Bell inequalities is optimally violated and used to certify the randomness of the outcomes.
Hence, measured data from Alice and ${\rm Bob}_k$ are post-selected according to the prepared state $\ket{\psi_{k,\vec h}}$.
It is important to underline that it is always possible for Alice and ${\rm Bob}_k$'s to choose their bases outside of the light cones of the outcomes obtained by ${\rm Bob}_1,\cdots,{\rm Bob}_{k-1}$, physically enforcing independence between the inputs of the Bell test and the post-selection.
In this way, dishonest devices cannot influence the outcomes of the test by exploiting the post-selection.

Finally, regardless of whether she takes a guess or not, Alice needs to be able to perform an exponentially growing number of different unitary transformations, which is a further practical difficulty.

\subsection{Nonmaximal violations}
\label{sub-sec:non_maximal_violations}
The authors of Ref. \cite{Curchod2017a} also conjecture and numerically verify a relation that bounds the guessing probability $G_k$ of the outcomes of $B^{(1)}_k$ with a nonmaximal violation of the Bell inequality:
\begin{equation}
    G_k \leq G_{\mathrm{max},k} = 	\frac{1}{2} + \frac{\sqrt{I_{\mathrm{max},k}^2 - I_k^2}}{2\qty(2-\beta_k)}.
\label{eq:guessing}
\end{equation}
This is important not only because experimental imperfections make maximal violations effectively impossible to observe, but also because the protocol itself requires $\xi_k >0$ for all but at most the very last step, which means that $I_k = I_{\mathrm{max},k}$ would be unattainable even if a perfect apparatus were used.

This means that after observing $I_k$, Alice and ${\rm Bob}_k$ can conclude that the min-entropy of each outcome of $B^{(1)}_k$ is $H_{\mathrm{min},k} = -\log_2{G_{\mathrm{max},k}}$, with $G_{\mathrm{max},k}$ calculated as in Eq. \eqref{eq:guessing}.
Therefore, at any step $k$ a close-to-maximal violation of the Bell inequality allows to extract close-to-1 random bits from each outcome.
The outcomes of $B^{(0)}$ do not contribute to this extraction, but the performance loss can be minimized by choosing $B^{(1)}$ with high probability.

\subsection{Robustness to imperfections}
\label{sub-sec:robustness}    
To account for real-world imperfections, we consider an initial state described by the density matrix 
\begin{equation}
    \label{eq:state_model_vis}
    \rho_1 = \qty(1-p-c)\dyad{\psi_1} + p \frac{\mathbb{1}}{4} + c \frac{\dyad{00} + \dyad{11}}{2},
\end{equation}
where we are setting $\theta_1 = \frac{\pi}{4}$ in the definition of $\ket{\psi_1}$ to use the best possible state as a starting point.
The second addend introduces diagonal terms in the ideal density matrix so that the state becomes depolarized.
It models the mixing of the ideal state with uncorrelated noise, such as, in the case of a photonics-based experiment, background light, dark counts or accidental coincidences.
The third addend induces decoherence in the state because the extreme antidiagonal terms of the matrix are reduced with respect to the diagonal ones.
It is especially realistic for states produced via SPDC, for which the indistinguishability between the $\ket{00}$ and $\ket{11}$ components is the result of precise alignment.
Unavoidable small inaccuracies generate the classical superpositions described by this addend.
This simple model is convenient because parameters $p$ and $c$ are easy to estimate experimentally.
Indeed, they are directly related to the visibilities $V_{\mathcal{Z}}$ and $V_{\mathcal{X}}$ of the state:
\begin{equation}
\label{eq:VisibilityPC}
    \begin{aligned}
        V_{\mathcal{Z}} &= \Tr(\sigma_Z \otimes \sigma_Z \rho_1) = 1-p, \\
        V_{\mathcal{X}} &= \Tr(\sigma_X \otimes \sigma_X \rho_1) = 1-p-c,
    \end{aligned}
\end{equation}
where we are assuming $V_{\mathcal{Z}} \geq V_{\mathcal{X}}$.
This is common in sources of polarization-entangled photon pairs, whose polarizing elements define a privileged basis, usually labeled $\mathcal{Z}$, for which visibility is higher.
Visibilities are a straightforward characterization technique for such sources, and allow us to easily calculate $p$ and $c$ and to compare the experimental results with the theoretical predictions.

By applying the protocol described in Sec. \ref{sec:model} to the initial state of Eq. \eqref{eq:state_model_vis}, we can evaluate how robust the results can be to imperfections in the preparation of the entangled pair.
For simplicity, we set $c=0$ in this initial characterization, and we will use the full model compared to the experimental results in Sec. \ref{sec:results}.

\begin{figure}
    \centering
    \includegraphics[width=\linewidth]{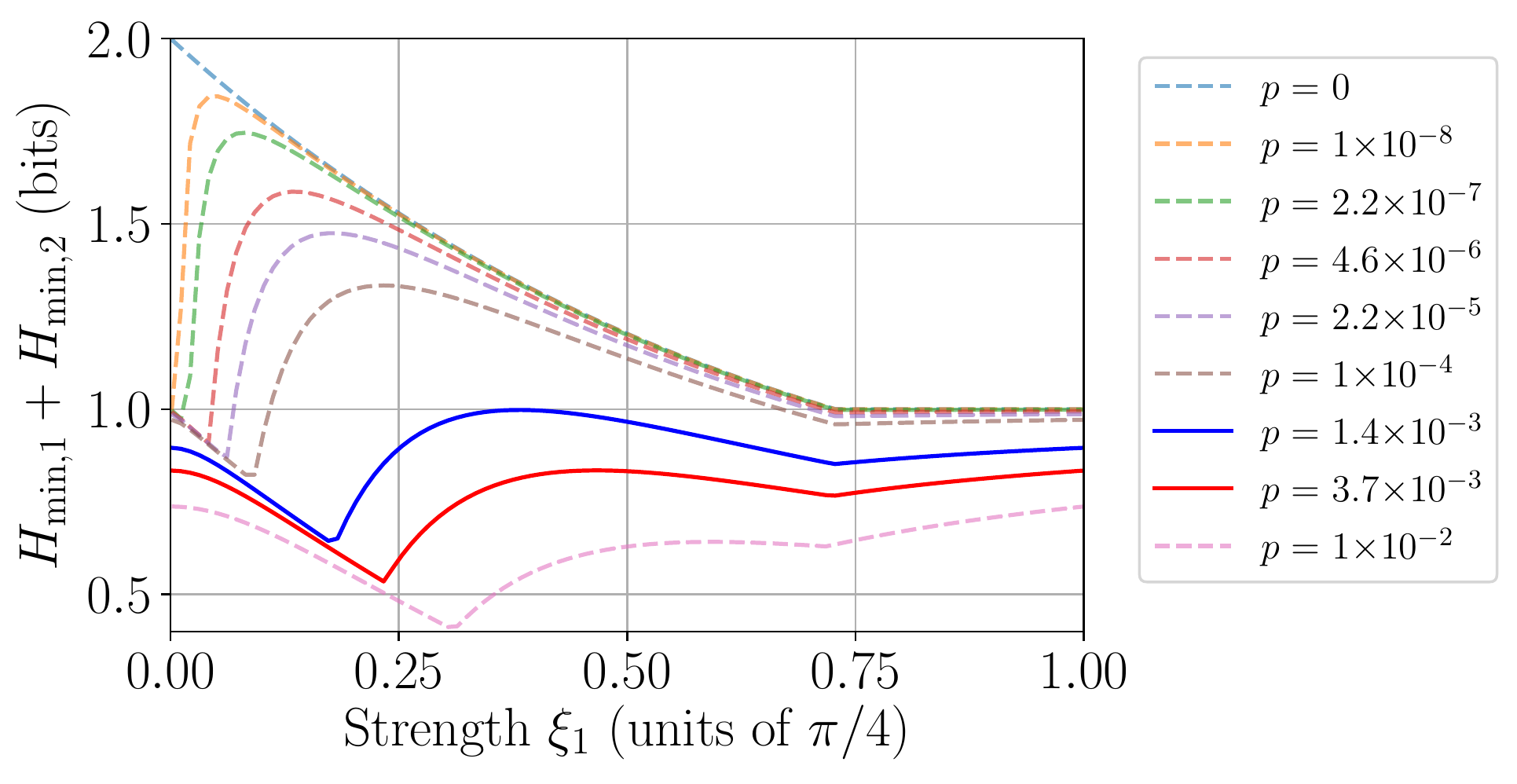}\\
    \caption{Achievable secure bits from one weak extraction and a subsequent projective one, for several values of the parameter $p$. The two highlighted solid lines are related to the values of $p$ that allow to reach $1$ total secure bit (blue) and for which one projective measurement starts to outperform the two-steps protocol ($p_{thr}^{(12)}$, red).}
    \label{fig:H1plusH2_charact}
\end{figure}

In Fig. \ref{fig:H1plusH2_charact}, we show the min-entropy $H_{\mathrm{min}}$ for a sequence of only two steps, with the last being projective, for several values of $p$.
The curves, which show the sum of the contributions of the two steps, have two kinks and therefore can be divided in three regions.
In the leftmost one, the first measurement is too strong to preserve entanglement, therefore $H_{\mathrm{min}, 2}=0$ and only the first step contributes.
The opposite happens in the rightmost region, in which $H_{\mathrm{min}, 1}=0$.
The region between the two kinks is the most interesting, because here both measurement steps contribute to the production of random bits.
For $p<p_{thr}^{(12)} \approx 3.7\cdot 10^{-3}$ the global maximum of the curve is achieved inside this region, indicating that a weak measurement is optimal.
For instance, for $p= 1.4 \cdot 10^{-3}$, the protocol can achieve about $1$ bit of min-entropy from the sum of the two steps, using $\xi_1 \approx 0.3$, whereas a single projective measurement would reach only $0.9$ bits.
As predicted in Ref. \cite{Curchod2017a}, the optimal value of $\xi_1$ is very close to 0 for nearly ideal states, but grows with the introduction of depolarization.
For $p>p_{thr}^{(12)}$, this value is 0, indicating that the best strategy is to use a single projective measurement.

\begin{figure}
    \centering
    \includegraphics[width=\linewidth]{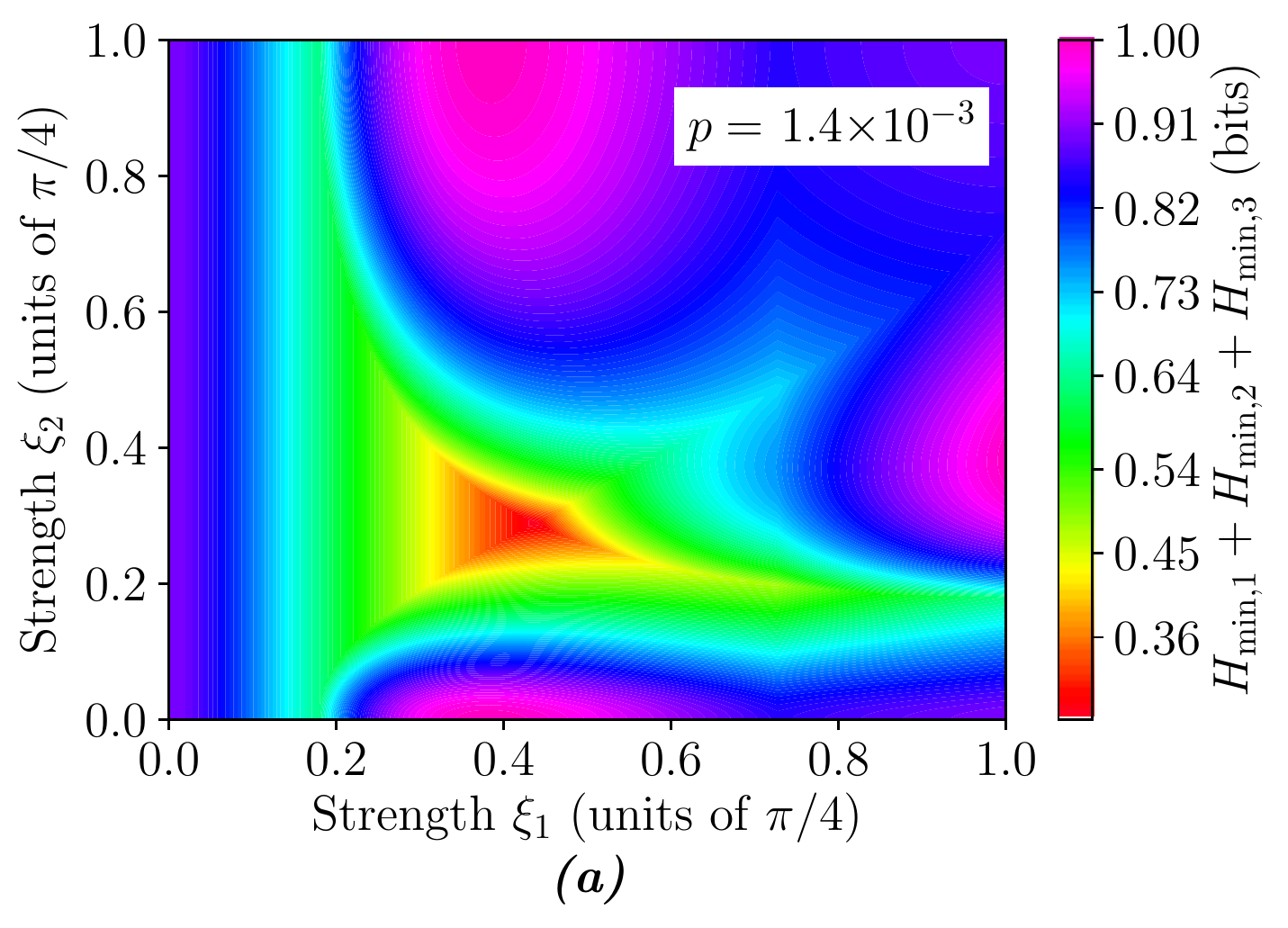}\\
    \includegraphics[width=\linewidth]{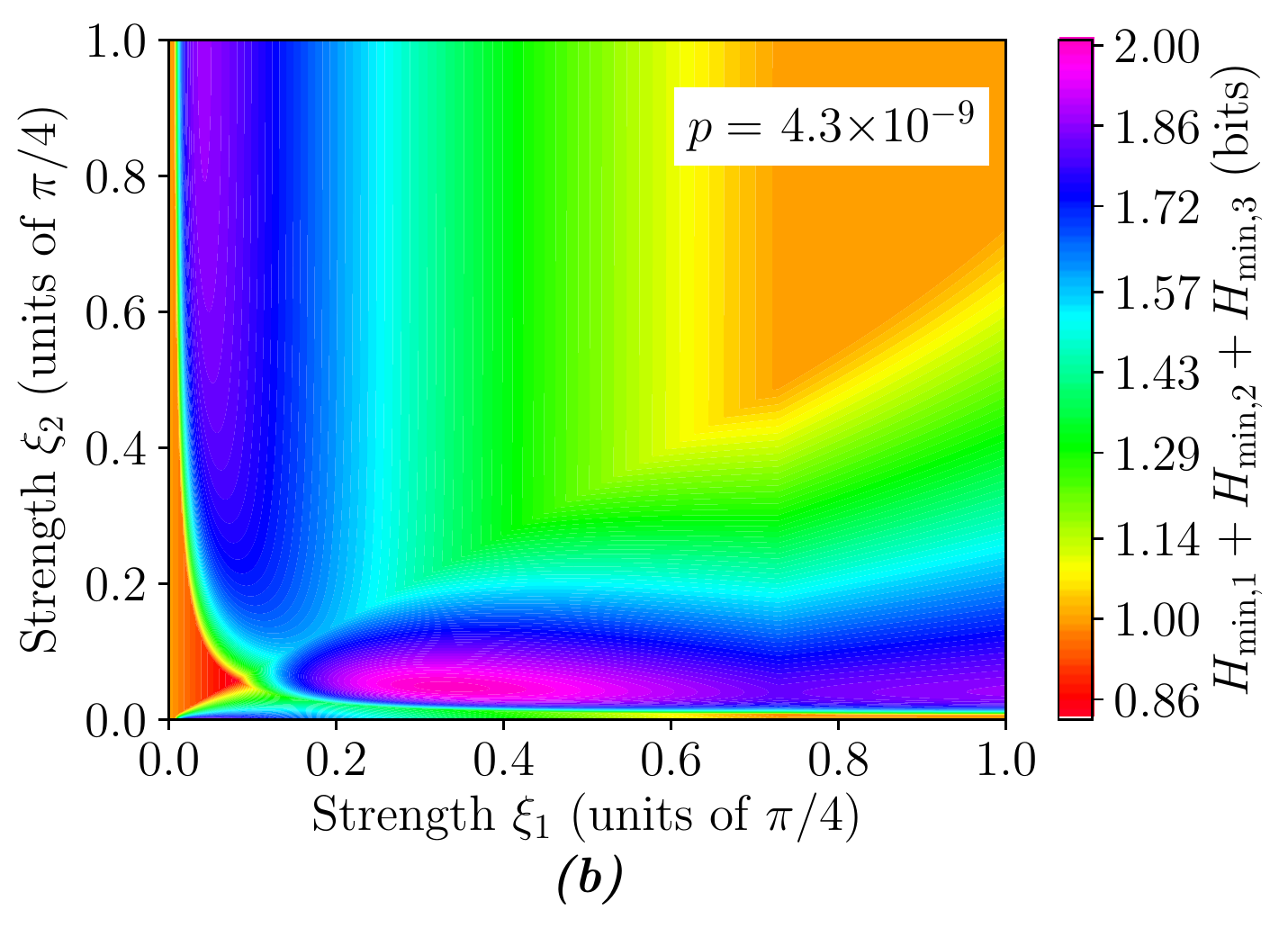}
    \caption{Secure bits achievable from three extractions with the last one projective ($\xi_3=0$). With $p\approx 1.4 \cdot 10^{-3}$ it is possible to reach one bit only near the axes, i.e. when one of the three steps is not useful. For $p<4.3\cdot 10^{-9}$ more than two bits can be extracted.}
    \label{fig:H1plusH2plusH3_charact}
\end{figure}

We also evaluate the protocol with two weak steps and a third projective one.
From Fig. \ref{fig:H1plusH2plusH3_charact}(a), we can see that $p=1.4 \cdot 10^{-3}$ is too large to reach one bit except near the axes of the graph, i.e. when only two extractions are meaningful.
We numerically verified that with $p\approx 3.2 \cdot 10^{-4}$ it is possible to reach one bit with three nonzero extractions: this means that adding a third step to the protocol only worsens its robustness to depolarization.
Furthermore, in Fig. \ref{fig:H1plusH2plusH3_charact}(b) we also show the value of $p$ needed to reach two bits ($\approx 4.3\cdot 10^{-9}$).

\begin{figure}
    \centering
    \includegraphics[width=\linewidth]{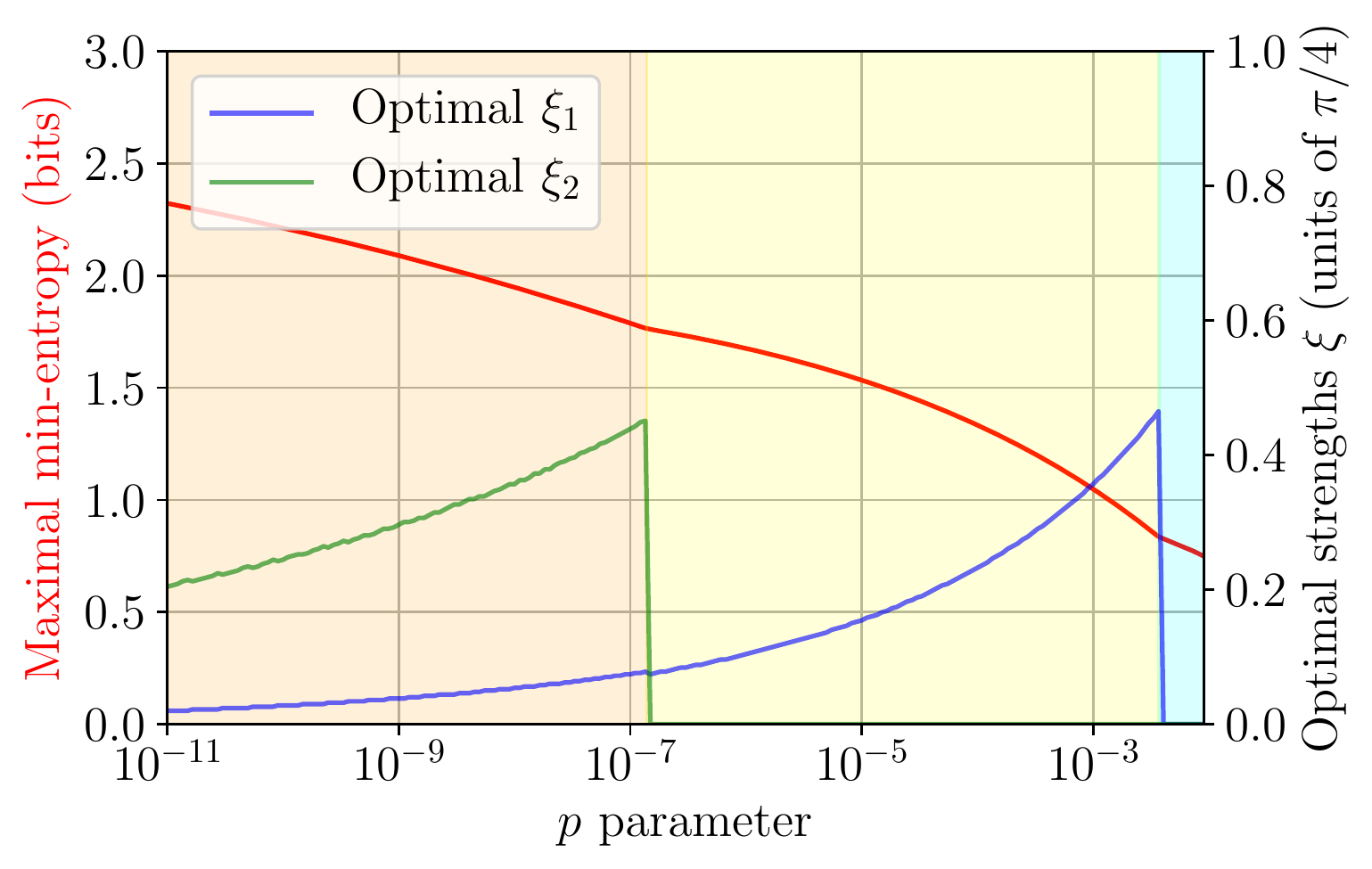}
    \caption{Maximal secure bits achievable as a function of the value of $p$. The colored regions represent the three cases in which it is better to perform one (cyan), two (yellow) or three (red) steps in order to achieve the maximal extraction of bits. The two threshold values are $p_{thr}^{(12)}\approx 3.7 \cdot 10^{-3} $ and $p_{thr}^{(23)} \approx 1.39\cdot 10^{-7}$.}
    \label{fig:max_bits_from_three_steps_charact}
\end{figure}

Finally, we investigate whether incrementing the sequence of measurements can increase the amount of extractable randomness.
Our analysis considers at most three steps (the last of which projective) and consists of a numerical maximization of the achievable min-entropy over the strength parameters $\xi_1$ and $\xi_2$ for each value of $p$.
We show the results in Fig. \ref{fig:max_bits_from_three_steps_charact}, where we highlight three possible cases with three different colors.
In the rightmost region (cyan), for which $p > p_{thr}^{(12)}$, the maximal extraction is achievable by setting $\xi_1=0$, i.e. with just one projective measurement.
By continuing to the left we find the interval for which $p_{thr}^{(23)} < p < p_{thr}^{(12)}$, where $p_{thr}^{(23)} \approx 1.39\cdot 10^{-7}$ (yellow).
Here, the strategy that maximizes randomness is to begin with a weak measurement ($\xi_1 \neq 0$), and then stop after a second projective one ($\xi_2 = 0$).
The leftmost region (red) indicates where two weak extractions plus a projective one outperform the previous two cases: indeed, the maximal min-entropy is achievable if $\xi_1 \neq 0$ and also $\xi_2 \neq 0$.
From these results, we can conclude that longer sequences are only beneficial for smaller and smaller amounts of depolarization.

\section{EXPERIMENTAL METHOD} 
\begin{figure}
    \centering
    \includegraphics[width=\linewidth]{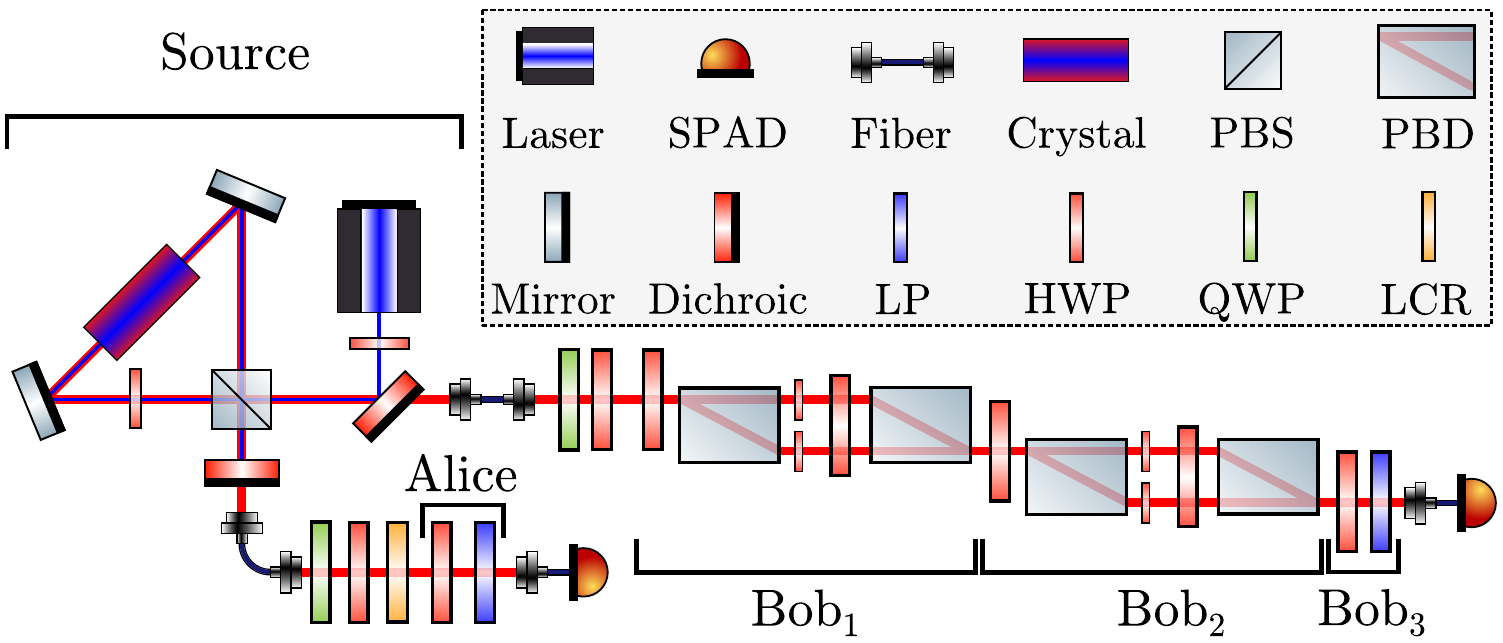}
    \caption{Scheme of the experimental setup.}
    \label{fig:setup_of_experiment}
\end{figure}

We verified experimentally whether these weak measurements could violate the Bell inequality strongly enough to generate randomness.

The source of polarization-entangled photons is a $30$ mm long periodically poled potassium titanyl phosphate (PPKTP) crystal, placed inside a Sagnac interferometer.
A continuous-wave laser at $404$ nm provides the pump light, that enters the polarizing beam splitter of the Sagnac interferometer with a diagonal polarization.
The exiting photons at $808$ nm are then collected into two single-mode fibers and brought to Alice and the Bobs' measurement sides.
Here, a half-wave plate (HWP) and a quarter-wave plate are used by both parties to remove the unitary evolutions due to the fibers and to transform the state into the desired one 
\begin{equation}
    \ket{\psi_1} = \frac{1}{\sqrt{2}}\qty(\ket{HH} + \ket{VV}),
\end{equation}
where the horizontal ($\ket{H}$) and vertical ($\ket{V}$) polarization components correspond to the $\ket{0}$ and $\ket{1}$ states of the theoretical protocol.
Furthermore, Alice uses a liquid-crystal retarder (LCR) to fine tune the phase between the two different polarization components.
Since Alice needs to measure only linear polarizations, her measurement setup consists of an HWP and a linear polarizer (LP). 
The complete scheme of the setup is depicted in Fig. \ref{fig:setup_of_experiment}.

On the Bobs' side we used a series of two Mach-Zehnder interferometers (MZI) that implement the weak measurements described in the protocol.
Each of them is composed by two polarizing beam displacers (PBD) that separate and rejoin the horizontal and vertical polarization components, two small HWPs (one per arm) and a shared HWP that selects the strength of the measurement.
Two more HWPs, one before and one after the MZI choose the basis for the measurement and apply the unitary operations $U^\dagger_{B}$.
The small HWP in the H path has its fast axis horizontal, while the one in the V path is rotated by $\pi/4$.
The strength $\xi$ of the measurement is regulated by setting the shared HWP at $\pi/4 - \xi/2$.

After the two MZIs, a third projective polarization measurement is implemented by an HWP and a LP.	
At each side, after the evolution of the state, the photons are collected into a fiber and sent to a single-photon avalanche diode connected to an $80$-ps resolution time tagger that returns coincidence counts within a $\pm 1$-ns window.

By rotating the HWPs between the interferometers, we set not only the measurement bases, but also the outcomes that correspond to the photons that pass through the only exit of the PBDs that is connected to the rest of the setup.
We can then scan all the different combinations of bases and outcomes sequentially, and, for each of them, record the number of coincident events.
The exposure time is fixed and chosen to gather enough statistics to obtain small statistical errors (details in Sec. \ref{sec:results}).
Then, linear combinations of the coincidence rates allow us to estimate the expected values in Eq. \eqref{eq:bellQuantity} and ultimately $I_k$.

In order to truly observe all outcomes without choosing them beforehand, as is necessary to produce random bits, this setup would require a treelike structure on Bob's side, which would grow exponentially with the number of steps.
We also note that the violations of the Bell inequality that we report are affected by several loopholes, such as the locality and detection ones \cite{Larsson2014}.
More profoundly, we do not choose the bases randomly, and do not record random outcomes from each measurement, but only expectation values, hence this is not a true Bell experiment.
A faithful implementation of the protocol should address all these issues.
However, our setup allows the feasibility study of the weak measurements, which is the focus of this work.

\section{RESULTS}
\label{sec:results}

\begin{figure}
    \centering
    \includegraphics[width=\linewidth]{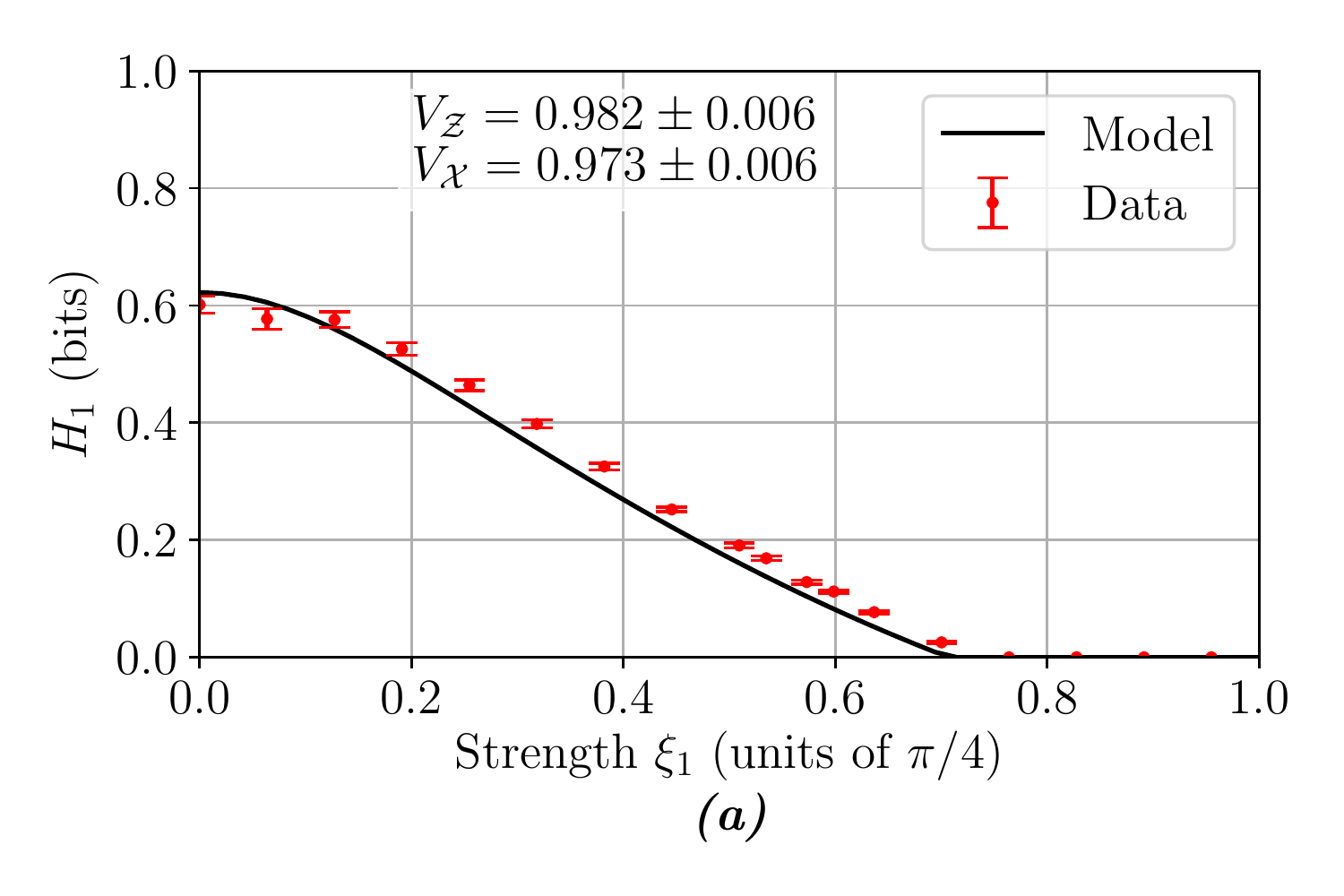}\\
    \includegraphics[width=\linewidth]{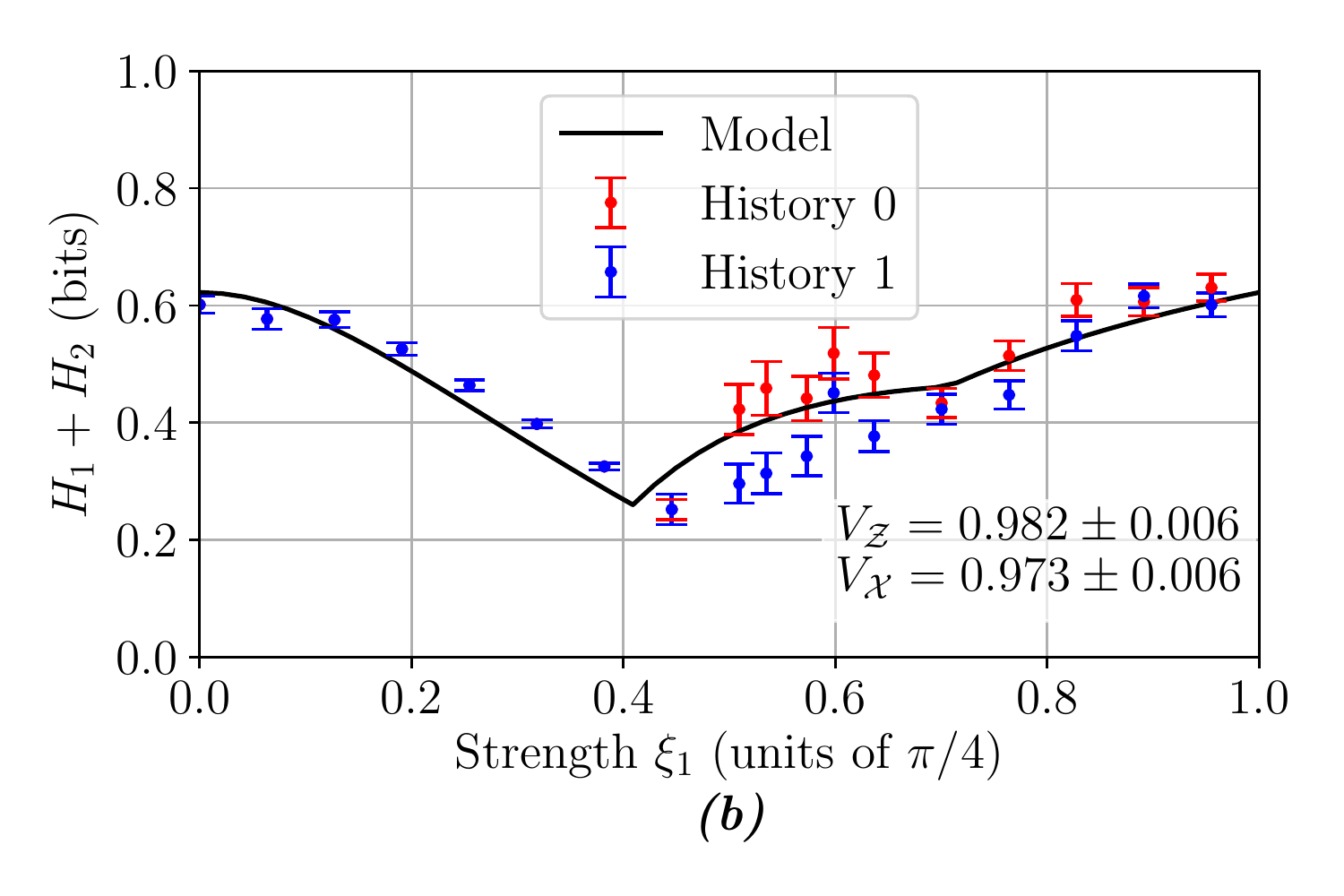}\\
    \includegraphics[width=\linewidth]{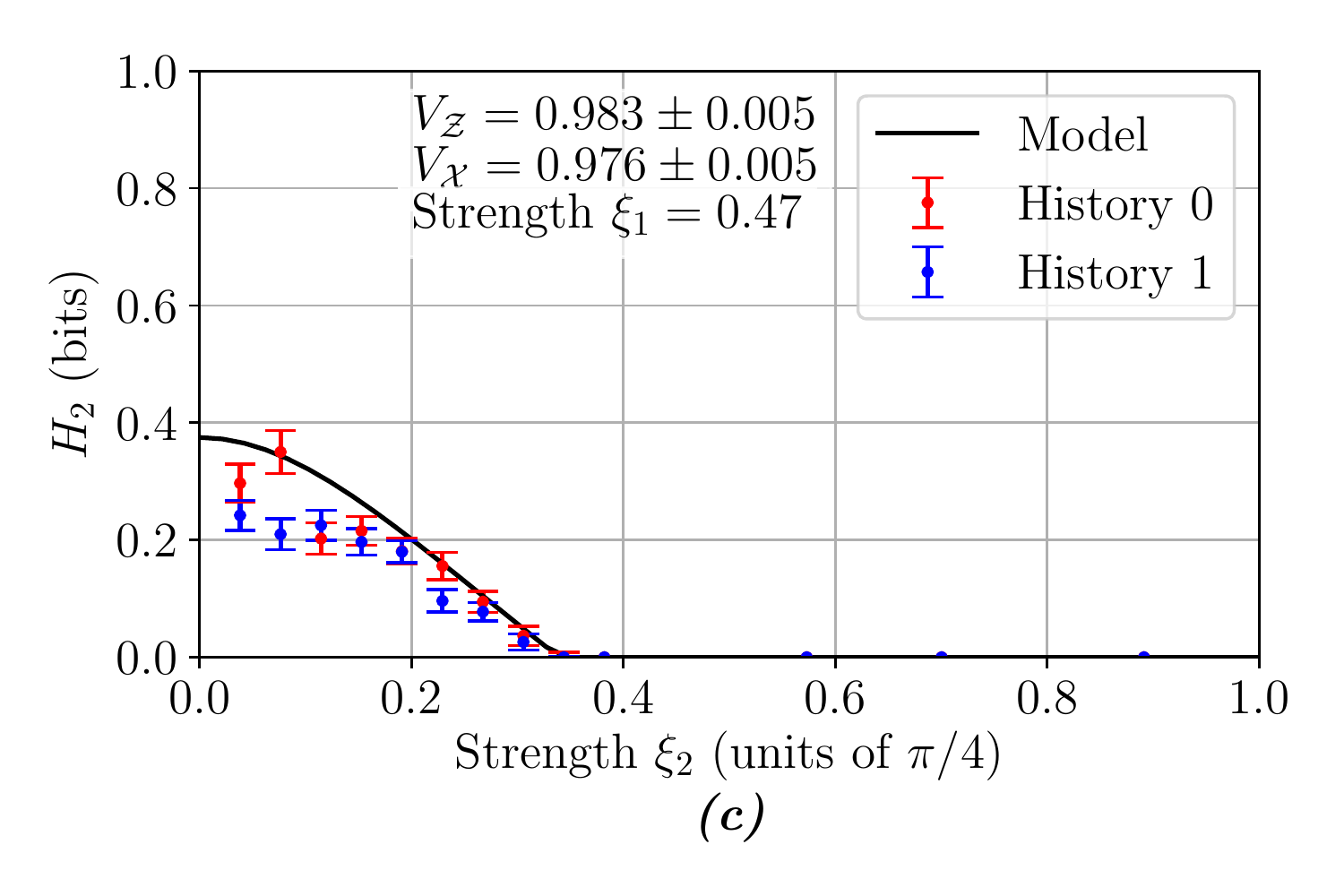}
    \caption{Results of the feasibility tests with variable strength. The continuous line is the model prediction while the dots represent the experimental data with their standard deviations, calculated with a Monte Carlo simulation which considers the Poissonian error on photon counting.}
    \label{fig:scan_result}
\end{figure}

We first characterize the protocol as a function of the strength parameters, adopting the model of Eq. (\ref{eq:state_model_vis}) for the initial state.
After finding the best strength, we perform longer-exposure experiments with that setting, like a real implementation of a randomness generator would do.
We choose an exposure time of about $60$s ($\sim 2\cdot10^5$ detected coincidences) for the long-exposure tests and about $30$s ($\sim 10^5$ detected coincidences) for the variable-strength tests.
This makes statistical errors small enough.
We do not consider any finite-statistics effect in the estimation of the min-entropy.
As we can see from Sec. \ref{sub-sec:robustness}, the protocol is very demanding in terms of purity of the state and, with our experimental visibilities, we must stop at two extraction steps, since we cannot violate the Bell inequality afterward.

Figure \ref{fig:scan_result}(a) shows the number of secure bits achievable from one weak step at different values of $\xi_1$ compared with the prediction of our model calculated with the initial visibilities reported in the chart.
In Fig. \ref{fig:scan_result}(b) we can see the previous bits summed with the bits extracted from a subsequent projective step, while Fig. \ref{fig:scan_result}(c) shows the result of a second weak step after a first one with $\xi_1 = 0.47$.

In order to estimate the quality of our state, we measure the initial visibilities in bases $\mathcal{X} = \{\ket{D},\ket{A}\}$ and $\mathcal{Z} = \{\ket{H},\ket{V}\}$ before the experiments.
To do this, we use an half-wave plate and a linear polarizer in front of the SPAD in the same way that we perform projective measurements for the protocol.
We calculate their standard deviations via propagation assuming Poissonian counting errors at the detectors.
We report these values in Fig. \ref{fig:scan_result}.
From them, we can use Eq. \eqref{eq:VisibilityPC} to obtain parameters $p$ and $c$, which we insert in the model to predict the amount of extractable randomness.

We can clearly see that our imperfect preparation prevents us from generating more than one bit of randomness per entangled pair, and enlarges the region of strength parameters where we cannot violate the Bell inequality at all.
Although we can generate some randomness at both the first and the second step, we achieve the best results when one measurement is noninteractive and the other is projective: Our state is too depolarized to make the weak measurement useful. 
Yet, our results closely follow the theoretical predictions, especially at the first step.
The second interferometer, by introducing further imperfections in the measurement, makes our data slightly separate from the solid line, as seen in Fig. \ref{fig:scan_result}(c).

Table \ref{tab:feasibility_tests_results} shows the results of the long-exposure feasibility tests compared with the model prediction.
We choose a strength $\xi_1 = 0.4$ when the next step is projective, while we choose $\xi_1 = 0.47$ and $\xi_1 = 0.52$ in order to perform a second nonprojective step with $\xi_2 = 0.1$.
The two rightmost columns show the min-entropy predicted by the model and measured experimentally: Our results are slightly below the predictions, probably because of systematic misalignments in the optical setup.

Albeit not shown in the table, we also add a third projective step after the second weak one, but the correlations between Alice and ${\rm Bob}_3$'s results are not strong enough to violate the CHSH-like inequality, and hence provide $H_{\mathrm{min},3} = 0$.
We attribute this to the visibilities of the state we produced, which do not allow more than two extractions of randomness, as predicted by our analysis. 

\begin{table*}
    \caption{Results of the long-exposure feasibility tests.
    Standard deviations are calculated with a Monte Carlo simulation which considers the Poissonian error on photon counting.}
    \begin{tabular*}{\linewidth}{l @{\extracolsep{\fill}}cccc}
        \toprule \toprule
         & & Strength & $H_{\mathrm{min},k}$ (Model) & $H_{\mathrm{min},k}$ (Experiment)\\
        Step $k$ & Previous outcome & (rad) & (bits) & (bits)\\\midrule
        $1$ & Not applicable & $0.4$ & $0.165$ & $0.13 \pm 0.002$\\
        $2$ & $0$ & Projective & $0.263$ & $0.38 \pm 0.04$\\
        $2$ & $1$ & Projective & $0.263$ & $0.13 \pm 0.02$\vspace{1ex}\\
        $1$ & Not applicable & $0.47$ & $0.085$ & $0.057 \pm 0.002$\\
        $2$ & $0$ & $0.1$ & $0.303$ & $0.32 \pm 0.02$\\
        $2$ & $1$ & $0.1$ & $0.303$ & $0.25 \pm 0.02$\vspace{1ex}\\
        $1$ & Not applicable & $0.52$ & $0.035$ & $0.005 \pm 0.001$\\
        $2$ & $0$ & $0.1$ & $0.369$ & $0.38 \pm 0.02$\\
        $2$ & $1$ & $0.1$ & $0.369$ & $0.33 \pm 0.01$\\
        \bottomrule\bottomrule
        \label{tab:feasibility_tests_results}
    \end{tabular*}
\end{table*}

\section{CONCLUSIONS}
In this work, we have studied the feasibility of using sequential weak measurements to extract more randomness from entangled pairs.
We have evaluated the protocol of Ref. \cite{Curchod2017a} and focused on its robustness to imperfections in the preparation of the initial quantum state.
Our analysis shows that even small amounts of depolarization nullify the performance gain (in terms of produced random bits per entangled pair) offered by the addition of a new measurement in the sequence, and the longer the sequence, the closer to ideal the state has to be in order to fully exploit all measurements.
For instance, a second step is useful only for $p < p_{thr}^{(12)}\approx 3.7 \cdot 10^{-3}$ and a third for $p<p_{thr}^{(23)} \approx 1.39\cdot 10^{-7}$.

Our experiment fully confirms the validity of this protocol and of the model summarized by Eq. \eqref{eq:state_model_vis}, which includes most of the inaccuracies of our setup, as can be seen by the resemblance between the data points and theoretical predictions in Fig. \ref{fig:scan_result}.
Moreover, it produces correlations that would allow to extract up to approximately $0.6$ bits of randomness from two sequential steps [Fig. \ref{fig:scan_result}(b)].
Yet, it further highlights the challenges in applying this protocol and even just in the preparation of an accurate enough entangled state.
Although there are reports of better visibilities \cite{Poh2015}, the unavoidable imperfections of bulk optical components make it difficult to reduce the value of $p$ much below $p_{thr}^{(12)}$.

However, a simple model such as ours indicates the minimum quality of the initial entangled state required to make the sequential protocol useful.
Through it, other experimental platforms could be investigated.
Integrated optics can offer polarizing beam splitters with comparable extinction ratio \cite{Zhang2016,Li2017,Ong2017}, and although entanglement sources do not yet reach the same visibilities, their quality \cite{Meyer-Scott2018} and capabilities \cite{Wang2018} are developing quickly.
In the field of quantum computing, two-qubit gates with fidelities well above $99\%$ have been demonstrated \cite{Barends2014,Ballance2016}, and perhaps, with some more improvements, the same technologies could be used to produce entangled states of the necessary quality for protocols like this.
Alternatively, similar schemes that are more robust to noise could be considered.
For instance, the protocol of Ref. \cite{Bowles2020} uses a weak measurement followed by a three-outcome POVM.
Albeit limited to only two steps, it can overcome the bound of two bits generated from one-half of an entangled pair, even in experimentally viable noise conditions.
Techniques like this will probably allow sequential weak measurements to improve the performance of randomness extraction with presently available optical components.

\begin{acknowledgments}
    Part of this work was supported by Ministero dell'Istruzione, dell'Università e della Ricerca (MIUR) (Italian Ministry of Education, University and Research) under the initiative ``Departments of Excellence'' (Law No. 232/2016), and by Fondazione Cassa di Risparmio di Padova e Rovigo within the call ``Ricerca Scientifica di Eccellenza 2018'',  project \textit{QUASAR}.
\end{acknowledgments}

\bibliography{bibliography}

\end{document}